\pdfoutput=1
\documentclass{sig-alternate-05-2015}

\usepackage{amsmath}
\usepackage{bbm}
\usepackage{booktabs}
\usepackage[labelfont=bf,textfont=bf,skip=0pt]{caption}
\usepackage{color}
\usepackage{enumitem}
\usepackage{flushend}
\usepackage{footnote}
\usepackage{grffile}
\usepackage{multirow}
\usepackage[all]{nowidow}
\usepackage[np, autolanguage]{numprint}
\usepackage{setspace}
\usepackage{subcaption}
\usepackage{times}
\usepackage{tikz}
\usepackage{tikz-3dplot}
\usepackage{todonotes}
\usepackage[square,comma,numbers,sort&compress,sectionbib]{natbib}
\usepackage{paralist}
\usepackage{url}
\usepackage{xintfrac}

\def \paperImplementationUrl {\url{https://github.com/cvangysel/sesh}}

\allowdisplaybreaks

\def \QCM {QCM}
\def \Nugget {Nugget}
\def \NuggetLexical {Nugget (RL2)}

\def \TF {TF}

\CopyrightYear{2016}
\setcopyright{acmlicensed}
\conferenceinfo{ICTIR '16,}{September 12 - 16, 2016, Newark, DE, USA}
\isbn{978-1-4503-4497-5/16/09}\acmPrice{\$15.00}
\doi{http://dx.doi.org/10.1145/2970398.2970422}

\def \NDCG {NDCG@10}
\def \MeanReciprocalRank {MRR}

\begin{document}

\title{Lexical Query Modeling in Session Search}

\numberofauthors{1}
\author{
\alignauthor
\mbox{}\hfill
\begin{tabular}{c}
    Christophe Van Gysel\\
   \email{cvangysel@uva.nl}
\end{tabular}
\hfill
\begin{tabular}{c}
    Evangelos Kanoulas\\
   \email{e.kanoulas@uva.nl}
\end{tabular}
\hfill
\begin{tabular}{c}
    Maarten de Rijke\\
   \email{derijke@uva.nl}
\end{tabular}
\hfill\mbox{}\\[1.1ex]
   \affaddr{University of Amsterdam, Amsterdam, The Netherlands}
}

\maketitle
\begin{abstract}
Lexical query modeling has been the leading paradigm for session search. In this paper, we analyze TREC session query logs and compare the performance of different lexical matching approaches for session search. Naive methods based on term frequency weighing perform on par with specialized session models. In addition, we investigate the viability of lexical query models in the setting of session search. We give important insights into the potential and limitations of lexical query modeling for session search and propose future directions for the field of session search.
\end{abstract}


\section{Introduction}

Many complex information seeking tasks, such as planning a trip or buying a car, cannot sufficiently be expressed in a single query \citep{Hassan2014}. These multi-faceted tasks are exploratory, comprehensive, survey-like or comparative in nature \citep{Raman2013} and require multiple search iterations to be adequately answered \citep{Kotov2011}. \citet{Donato2010} note that 10\% of the user sessions (more than 25\% of query volume) consists of such complex information needs.

The TREC Session Track \citep{TREC2011-2014} created an environment for researchers ``to test whether systems can improve their performance for a given query by using previous queries and user interactions with the retrieval system.'' The track's existence led to an increasing number of methods aimed at improving session search. \citet{Yang2015} introduce the Query Change Model (\QCM{}), which uses lexical editing changes between consecutive queries in addition to query terms occurring in previously retrieved documents, to improve session search. They heuristically construct a lexicon-based query model for every query in a session. Query models are then linearly combined for every document, based on query recency \citep{Yang2015} or document satisfaction \citep{Luo2014-TREC2014,TREC2014}, into a session-wide lexical query model. However, there has been a clear trend towards the use of supervised learning \citep{TREC2014,Yang2015,Luo2015} and external data sources \citep{Guan2012,Luo2014}. \citet{Guan2012} perform lexical query expansion by adding higher-order n-grams to queries by mining document snippets. In addition, they expand query representations by including anchor texts to previously top-ranked documents in the session. \citet{TREC2014} expand document representations by including incoming anchor texts. \citet{Luo2015} introduce a linear point-wise learning-to-rank model that predicts relevance given a document and query change features. They incorporate document-independent session features in their ranker.

The use of machine-learned ranking and the expansion of query and document representations is meant to address a specific instance of a wider problem in information retrieval, namely the query document mismatch \citep{Li2014}. In this paper, we analyze the session query logs made available by TREC and compare the performance of different lexical query modeling approaches for session search, taking into account session length.\footnote{An open-source implementation of our testbed for evaluating session search is available at \paperImplementationUrl{}.} In addition, we investigate the viability of lexical query models in a session search setting.

The main purpose of this paper is to investigate the potential of lexical methods in session search and provide foundations for future research. We ask the following questions:~%
\begin{inparaenum}[(1)]
	\item Increasingly complex methods for session search are being developed, but how do naive methods perform?
	\item How well can lexical methods perform?
	\item Can we solve the session search task using lexical matching only?
\end{inparaenum}

\section{Lexical matching for sessions}
\label{section:methodology}

\newcommand{\Prob}[1]{P(#1)}
\newcommand{\CondProb}[2]{\Prob{#1 \mid #2}}

\newcommand{\Vocabulary}{V}

\newcommand{\Sessions}{S}

\newcommand{\Corpus}{D}
\newcommand{\Session}{\MakeLowercase{\Sessions{}}}
\newcommand{\Queries}{Q}
\newcommand{\Terms}{T}
\newcommand{\SERPs}{R}

\newcommand{\Time}{i}

\newcommand{\Query}{\MakeLowercase{\Queries{}}}
\newcommand{\SERP}{\MakeLowercase{\SERPs{}}}
\newcommand{\Document}{\MakeLowercase{\Corpus{}}}

\newcommand{\Term}{\MakeLowercase{\Terms{}}}
\newcommand{\Word}{w}

\newcommand{\Length}[1]{|#1|}
\newcommand{\SERPLength}[1]{\Length{\SERP{}#1}}
\newcommand{\QueryLength}[1]{\Length{\Query{}#1}}

\newcommand{\NumInteractions}{n}
\newcommand{\TopK}{k}

\newcommand{\SessionModel}[1][\Session{}]{\theta^{#1}}
\newcommand{\SessionModelParameter}[2][\Session{}]{\theta^{#1}_{#2}}

\newcommand{\DocumentModel}[1][]{\theta^{\Document{}{#1}}}
\newcommand{\DocumentModelParameter}[2][]{\theta^{\Document{}{#1}}_{#2}}

We define a search session $\Session{}$ as a sequence of $\NumInteractions{}$ interactions $(\Query{}_i, \SERP{}_i)$ between user and search engine, where $\Query{}_i$ denotes a user-issued query consisting of $\QueryLength{_i}$ terms $\Term{}_{i,1}$, \ldots, $\Term{}_{i,{\QueryLength{_i}}}$ and $\SERP{}_i$ denotes a result page consisting of $\SERPLength{_i}$ documents $\SERP{}_{i, 1}$, \ldots, $\SERP{}_{i, \SERPLength{_i}}$ returned by the search engine (also referred to as SERP). The goal, then, is to return a SERP $\SERP{}_{\NumInteractions{} + 1}$ given a query $\Query{}_{\NumInteractions{} + 1}$ and the session history that maximizes the user's utility function.

In this work, we formalize session search by modeling an observed session $\Session{}$ as a query model parameterized by $\SessionModel{} = \{ \SessionModelParameter{1}$, \ldots, $\SessionModelParameter{\Length{\Vocabulary{}}} \}$, where $\SessionModelParameter{i}$ denotes the weight associated with term $\Term{}_i \in \Vocabulary{}$ (specified below). Documents $\Document{}_j$ are then ranked in decreasing order of
\begin{equation*}
\log\CondProb{\Document{}_j}{\Session{}} = \sum^{\Length{\Vocabulary{}}}_{k=1} \SessionModelParameter{k} \log\DocumentModelParameter[_j]{k},
\end{equation*}
where $\DocumentModel[_j]$ is a lexical model of document $\Document{}_j$, which can be a language model (LM), a vector space model or a specialized model using hand-engineered features. Query model $\SessionModel{}$ is a function of the query models of the interactions $\Time{}$ in the session, $\SessionModel[\Session{}_\Time{}]{}$ (e.g., for a uniform aggregation scheme, $\SessionModel{} = \sum_\Time{} \SessionModel[\Session{}_\Time{}]{}$). Existing session search methods \citep{Yang2015,Guan2012} can be expressed in this formalism as follows:
\newcommand{\BestDocument}[1][1]{\SERP{}_{\Time{} - 1, #1}}
\newcommand{\idf}[1]{\text{idf}({#1})}
\begin{description}[topsep=0pt,itemsep=0pt,parsep=0pt]
	\item[Term frequency (\TF{})] Terms in a query are weighted according to their frequency in the query (i.e., $\SessionModelParameter[\Session{}_\Time{}]{k}$ becomes the frequency of term $\Term{}_k$ in $\Query{}_\Time{}$). Queries $\Query{}_\Time{}$ that are part of the same session $\Session{}$ are then aggregated uniformly for a subset of queries. In this work, we consider the following subsets: the \emph{first query}, the \emph{last query} and the \emph{concatenation of all queries} in a session. Using the last query corresponds to the official baseline of the TREC Session track \citep{TREC2014}.

	\item[\Nugget{}]
	\Nugget{} \citep{Guan2012} is a method for effective structured query formulation for session search. Queries $\Query{}_i$, part of session $\Session{}$, are expanded using higher order n-grams occurring in both $\Query{}_i$ and snippets of the top-$k$ documents in the previous interaction, $\BestDocument[1]{}$, \ldots, $\BestDocument[k]{}$. This effectively expands the vocabulary by additionally considering n-grams next to unigram terms. The query models of individual queries in the session are then aggregated using one of the aggregation schemes.
	\Nugget{} is primarily targeted at resolving the query-document mismatch by incorporating structure and external data and does not model query transitions. The method can be extended to include external evidence by expanding $\SessionModel{}$ to include anchor texts pointing to (clicked) documents in previous SERPs{}.

	\item[Query Change Model (\QCM{})]
	QCM \citep{Yang2015} uses syntactic editing changes between consecutive queries in addition to query changes and previous SERPs to enhance session search.
	In QCM \citep[Section~6.3]{Yang2015}, document model $\DocumentModel{}$ is provided by a language model with Dirichlet smoothing and the query model at interaction $\Time{}$, $\SessionModel[\Session{}_\Time{}]$, in session $\Session{}$ is given by%
	\begin{equation*}
	\small
	{\SessionModelParameter[\Session{}_\Time{}]{k}} = \begin{cases}
	1 + \alpha (1 - \CondProb{\Term{}_k}{\BestDocument{}}), & \!\Term{}_k \in \Query{}_{\text{theme}} \\

	1 - \beta \CondProb{\Term{}_k}{\BestDocument{}}, & \!\Term{}_k \in + \Delta \Query{} \wedge \Term{}_k \in \BestDocument{} \\
	1 + \epsilon \, \idf{\Term{}_k}, & \!\Term{}_k \in + \Delta \Query{} \wedge \Term{}_k \notin \BestDocument{} \\

	- \delta \CondProb{\Term{}_k}{\BestDocument{}}, & \!\Term{}_k \in - \Delta \Query{},
	\end{cases}
	\end{equation*}
	where $\Query{}_\text{theme}$ are the session's theme terms, $+ \Delta \Query{}$ ($- \Delta \Query{}$, resp.) are the added (removed) terms, $\CondProb{\Term{}_k}{\BestDocument{}}$ denotes the probability of $\Term{}_k$ occurring in SAT clicks, $\idf{\Term{}_k}$ is the inverse document frequency of term $\Term{}_k$ and $\alpha$, $\beta$, $\epsilon$, $\delta$ are parameters. The $\SessionModel[\Session{}_\Time{}]$ are then aggregated into $\SessionModel{}$ using one of the aggregation schemes, such as the uniform aggregation scheme (i.e., the sum of the $\SessionModel[\Session{}_\Time{}]$).
\end{description}
In \S\ref{section:discussion}, we analyze the methods listed above in terms of their ability to handle sessions of different lengths and contextual history.


\section{Experiments}
\label{section:experiments}

\begin{table*}[t]
	\centering

	\caption{Overview of 2011, 2012, 2013 and 2014 TREC session tracks. For the 2014 track, we report the total number of sessions in addition to those sessions with judgments. We report the mean and standard deviation where appropriate; M denotes the median.\label{tbl:statistics}}

	\resizebox{\textwidth}{!}{\begin{tabular}{lcccc}

\toprule
& 2011 & 2012 & 2013 & 2014 \\

\midrule

\textbf{Sessions} \\

Sessions & \numprint{76}  & \numprint{98}  & \numprint{87}  & \numprint{100} (\numprint{1021} total) \\
Queries per session & \nprounddigits{2} \npdecimalsign{.} \numprint{3.6842} $\pm$ \numprint{1.7933}; M=\numprint{3.0000}  & \nprounddigits{2} \npdecimalsign{.} \numprint{3.0306} $\pm$ \numprint{1.5744}; M=\numprint{2.0000}  & \nprounddigits{2} \npdecimalsign{.} \numprint{5.0805} $\pm$ \numprint{3.5983}; M=\numprint{4.0000}  & \nprounddigits{2} \npdecimalsign{.} \numprint{4.3389} $\pm$ \numprint{2.2228}; M=\numprint{4.0000} \\
Unique terms per session & \nprounddigits{2} \npdecimalsign{.} \numprint{7.0132} $\pm$ \numprint{3.2827}; M=\numprint{6.5000}  & \nprounddigits{2} \npdecimalsign{.} \numprint{5.7551} $\pm$ \numprint{2.9522}; M=\numprint{5.0000}  & \nprounddigits{2} \npdecimalsign{.} \numprint{8.8621} $\pm$ \numprint{4.3791}; M=\numprint{8.0000}  & \nprounddigits{2} \npdecimalsign{.}\numprint{7.7855} $\pm$ \numprint{4.0786}; M=\numprint{7.0000} \\

\midrule

\textbf{Topics} \\

Session per topic & \nprounddigits{2} \npdecimalsign{.} \phantom{00}\numprint{1.2258} $\pm$ \phantom{00}\numprint{0.4551}; M=\phantom{00}\numprint{1.0000}  & \nprounddigits{2} \npdecimalsign{.} \phantom{00}\numprint{2.0417} $\pm$ \phantom{00}\numprint{0.9781}; M=\phantom{00}\numprint{2.0000}  & \nprounddigits{2} \npdecimalsign{.} \phantom{00}\numprint{2.1803} $\pm$ \phantom{00}\numprint{0.9323}; M=\phantom{00}\numprint{2.0000}  & \nprounddigits{2} \npdecimalsign{.} \phantom{0}\numprint{20.9500} $\pm$ \phantom{00}\numprint{4.8146}; M=\phantom{0}\numprint{21.0000} \\
Document judgments per topic & \nprounddigits{2} \npdecimalsign{.} \numprint{313.1129} $\pm$ \numprint{114.6277}; M=\numprint{292.0000}  & \nprounddigits{2} \npdecimalsign{.} \numprint{372.1042} $\pm$ \numprint{162.6306}; M=\numprint{336.5000}  & \nprounddigits{2} \npdecimalsign{.} \numprint{268.0000} $\pm$ \numprint{116.8587}; M=\numprint{247.0000}  & \nprounddigits{2} \npdecimalsign{.} \numprint{332.3333} $\pm$ \numprint{149.0269}; M=\numprint{322.0000} \\

\midrule

\textbf{Collection} \\

Documents & \multicolumn{2}{c}{\numprint{21258800}} & \multicolumn{2}{c}{\numprint{15702181}} \\
Document length & \multicolumn{2}{c}{\nprounddigits{2} \npdecimalsign{.} \numprint{1096.18328043} $\pm$ \numprint{1502.45255801}} & \multicolumn{2}{c}{\nprounddigits{2} \npdecimalsign{.} \numprint{649.074948569} $\pm$ \numprint{1635.29462878}} \\
Terms & \multicolumn{2}{c}{\numprint{\xintFloat [3]{34015925}} (\numprint{\xintFloat [3]{23303541122}} total)} & \multicolumn{2}{c}{\numprint{\xintFloat [3]{23575957}} (\numprint{\xintFloat [3]{10191892325}} total)} \\
Spam scores & \multicolumn{2}{c}{GroupX} & \multicolumn{2}{c}{Fusion} \\

\bottomrule

\end{tabular}}

\end{table*}

\subsection{Benchmarks}

We evaluate the lexical query modeling methods listed in \S\ref{section:methodology} on the session search task (G1) of the TREC Session track from 2011 to 2014 \citep{TREC2011-2014}. We report performance on each track edition independently and on the track aggregate. Given a query, the task is to improve retrieval performance by using previous queries and user interactions with the retrieval system. To accomplish this, we first retrieve the \numprint{2000} most relevant documents for the given query and then re-rank these documents using the methods described in \S\ref{section:methodology}. We use the ``Category B'' subsets of ClueWeb09 (2011/2012) and ClueWeb12 (2013/2014) as document collections. Both collections consist of approximately 50 million documents. Spam documents are removed before indexing by filtering out documents with scores (GroupX and Fusion, respectively) below 70 \citep{Cormack2012}. Table~\ref{tbl:statistics} shows an overview of the benchmarks and document collections.

\subsection{Evaluation measures}

To measure retrieval effectiveness, we report Normalized Discounted Cumulative Gain at rank 10 (\NDCG{}) in addition to Mean Reciprocal Rank (\MeanReciprocalRank{}). The relevance judgments of the tracks were converted from topic-centric to session-centric according to the mappings provided by the track organizers.\footnote{We take into account the mapping between judgments and actual relevance grades for the 2012 edition.} Evaluation measures are then computed using TREC's official evaluation tool, {\tt trec\_eval}.\footnote{\url{https://github.com/usnistgov/trec_eval}}

\subsection{Systems under comparison}
\label{section:experiments:methods}

We compare the lexical query model methods outlined in \S\ref{section:methodology}. All methods compute weights for lexical entities (e.g., unigram terms) on a per-session basis, construct a structured Indri query \citep{Metzler2004} and query the document collection using {\tt pyndri}.\footnote{\url{https://github.com/cvangysel/pyndri}} For fair comparison, we use Indri's default smoothing configuration (i.e., Dirichlet smoothing with $\mu = 2500$) and uniform query aggregation for all methods (different from the smoothing used for \QCM{} in \citep{Yang2015}). This allows us to separate query aggregation techniques from query modeling approaches in the case of session search.

For \Nugget{}, we use the default parameter configuration ($k_\text{snippet}=10,\, \theta=0.97,\, k_\text{anchor}=5$ and $\beta=0.1$), using the strict expansion method. We report the performance of \Nugget{} without the use of external resources (RL2), with anchor texts (RL3) and with click data (RL4). For \QCM{}, we use the parameter configuration as described in \citep{Yang2015,Luo2015}: $\alpha=2.2,\, \beta=1.8,\, \epsilon=0.07$ and $\delta=0.4$.

In addition to the methods above, we report the performance of an oracle that always ranks in decreasing order of ground-truth relevance. This oracle will give us an upper-bound on the achievable ranking performance.

\subsection{Ideal lexical term weighting}
\label{section:experiments:ideal}

We investigate the maximally achievable performance by weighting query terms. Inspired by \citet{Bendersky2012}, we optimize \NDCG{} for every session using a grid search over the term weight space. We sweep the weight of every term between $-1.0$ and $1.0$ (inclusive) with increments of $0.1$, resulting in a total of $21$ weight assignments per term. Due to the exponential time complexity of the grid search, we limit our analysis to the \numprint{230} sessions with $7$ unique query terms or less (see Table~\ref{tbl:statistics}). This experiment will tell us the maximally achievable retrieval performance in session search by the re-weighting lexical terms only.


\section{Results \& Discussion}
\label{section:discussion}

\begin{table*}[t]
	\centering

	\caption{Overview of experimental results on 2011--2014 TREC Session tracks of the \TF{}, \Nugget{} and \QCM{} methods (see \S\ref{section:methodology}). The ground-truth oracle shows the ideal performance (\S\ref{section:experiments:methods}).\label{tbl:results}}

	{\begin{tabular}{c@{ }c@{ }c@{ }c@{ }c@{ }c@{ }c@{ }c@{ }c}
\toprule
\multirow{2}{*}{} & \multicolumn{2}{c}{2011} & \multicolumn{2}{c}{2012} & \multicolumn{2}{c}{2013} & \multicolumn{2}{c}{2014} \\ 
& NDCG@10 & MRR & NDCG@10 & MRR & NDCG@10 & MRR & NDCG@10 & MRR \\ 
\cmidrule(lr){2-3}
\cmidrule(lr){4-5}
\cmidrule(lr){6-7}
\cmidrule(lr){8-9}
\multicolumn{1}{l}{Ground-truth oracle} & $0.777$ & $0.868$ & $0.695$ & $0.865$ & $0.517$ & $0.920$ & $0.410$ & $0.800$ \\ 
\multicolumn{1}{l}{TF (first query)} & $0.371$ & $0.568$ & $0.302$ & $0.523$ & $0.121$ & $0.379$ & $0.120$ & $0.336$ \\ 
\multicolumn{1}{l}{TF (last query)} & $0.358$ & $0.598$ & $0.316$ & $0.586$ & $0.133$ & $0.358$ & $0.156$ & $0.458$ \\ 
\multicolumn{1}{l}{TF (all queries)} & $\phantom{}\textbf{0.448}$ & $\phantom{}\textbf{0.685}$ & $0.348$ & $0.604$ & $0.162$ & $0.477$ & $\phantom{}\textbf{0.174}$ & $\phantom{}\textbf{0.478}$ \\ 
\multicolumn{1}{l}{Nugget (RL2)} & $0.437$ & $0.677$ & $0.352$ & $0.609$ & $\phantom{}\textbf{0.163}$ & $0.488$ & $0.173$ & $0.476$ \\ 
\multicolumn{1}{l}{Nugget (RL3)} & $0.442$ & $0.678$ & $\phantom{}\textbf{0.360}$ & $\phantom{}\textbf{0.619}$ & $0.162$ & $\phantom{}\textbf{0.488}$ & $0.172$ & $0.477$ \\ 
\multicolumn{1}{l}{Nugget (RL4)} & $0.437$ & $0.677$ & $0.352$ & $0.609$ & $0.163$ & $0.488$ & $0.173$ & $0.476$ \\ 
\multicolumn{1}{l}{QCM} & $0.440$ & $0.661$ & $0.342$ & $0.575$ & $0.160$ & $0.484$ & $0.162$ & $0.450$ \\ 
\bottomrule
\end{tabular}
}

\vspace*{-0.75\baselineskip}
\end{table*}

\begin{figure}[t]

\newcommand{\boxplotinner}[2][]{%
	\IfFileExists{#2}{
		\includegraphics[width=\columnwidth#1]{#2}}{
		\resizebox{\columnwidth}{!}{\missingfigure{#2}}}%
}%

\boxplotinner{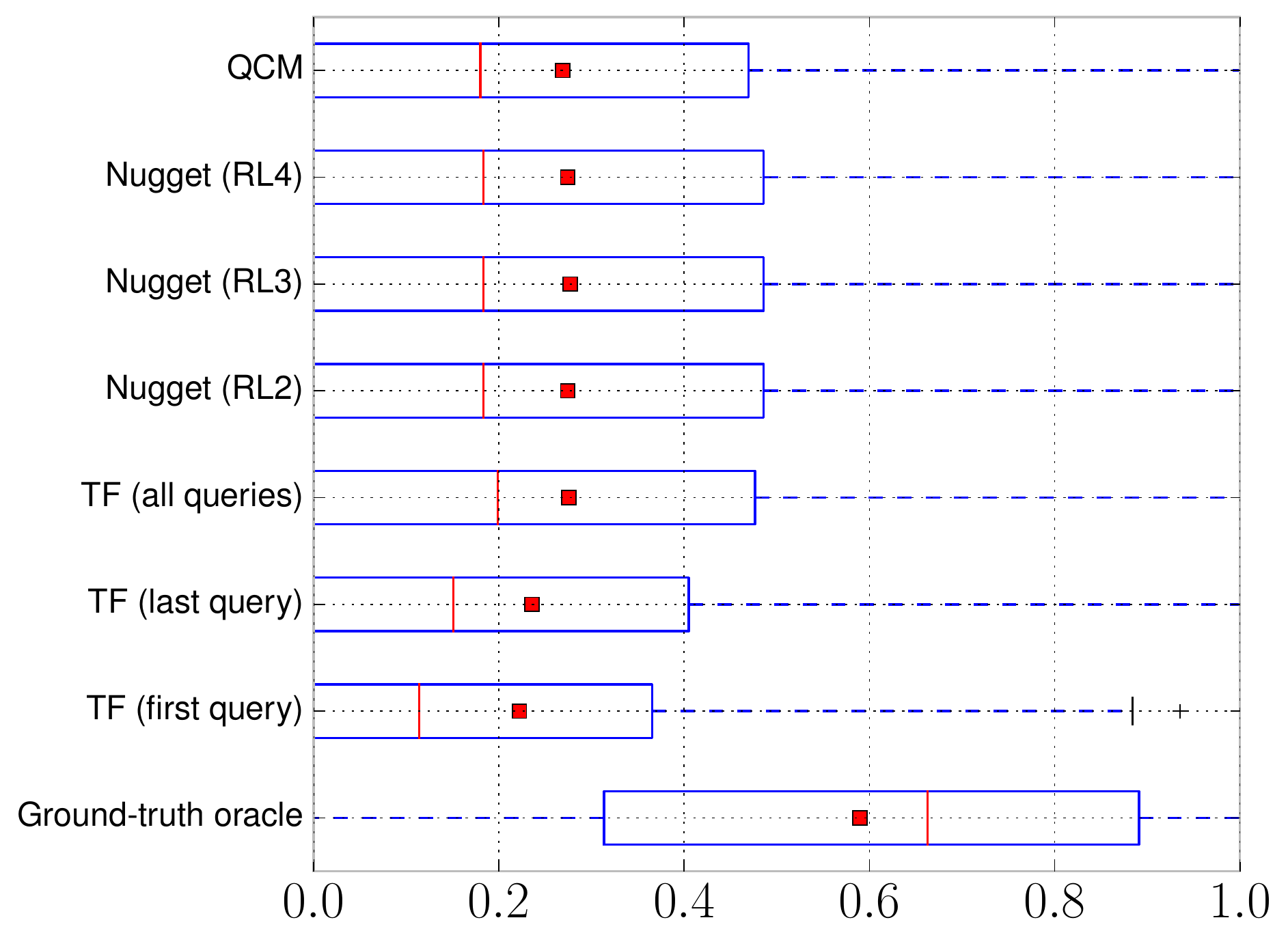}

\caption{Box plot of \NDCG{} on all sessions of the TREC Session track (2011--2014). The box depicts the first, second (median) and third quartiles. The whiskers are located at 1.5 times the interquartile range on both sides of the box. The square and crosses depict the average and outliers respectively.\label{fig:boxplot}}

\vspace*{-0.75\baselineskip}
\end{figure}

In this section, we report and discuss our experimental results. Of special interest to us are the methods that perform lexical matching based on a user's queries in a single session: \QCM{}, \NuggetLexical{} and the three variants of \TF{}.
Table~\ref{tbl:results} shows the methods' performance on the TREC Session track editions from 2011 to 2014. No single method consistently outperforms the other methods. Interestingly enough, the methods based on term frequency (\TF{}) perform quite competitively compared to the specialized session search methods (\Nugget{} and \QCM{}). In addition, the \TF{} variant using all queries in a session even outperforms \NuggetLexical{} on the 2011 and 2014 editions and \QCM{} on nearly all editions. Using the concatenation of all queries in a session, while being an obvious baseline, has not received much attention in recent literature or by TREC \citep{TREC2011-2014}. In addition, note that the best-performing (unsupervised) \TF{} method achieves better results than the supervised method of \citet{Luo2015} on the 2012 and 2013 tracks. Fig.~\ref{fig:boxplot} depicts the boxplot of the \NDCG{} distribution over all track editions (2011--2014). The term frequency approach using all queries achieves the highest mean/median overall. Given this peculiar finding, where a generic retrieval model performs better than specialized session search models, we continue with an analysis of the TREC Session search logs.

\begin{figure*}[t]

\newcommand{\lengthplotinner}[2][]{%
	\IfFileExists{#2}{
		\includegraphics[height=0.185\textheight#1]{#2}}{
		\resizebox{0.245\textwidth}{!}{\missingfigure{#2}}}%
}%
\newcommand{\lengthplot}[1]{%
	\def \PlotPath {resources/per_session_length/#1--exclude_method_names.pdf}%
	\begin{subfigure}[t]{0.225\textwidth}%
	\lengthplotinner{\PlotPath}%
	\caption{#1\label{fig:session_length:#1}}%
	\end{subfigure}%
}
\mbox{
\begin{subfigure}[t]{0.10\textwidth}%
\lengthplotinner[,trim=0 0 9.525cm 0, clip]{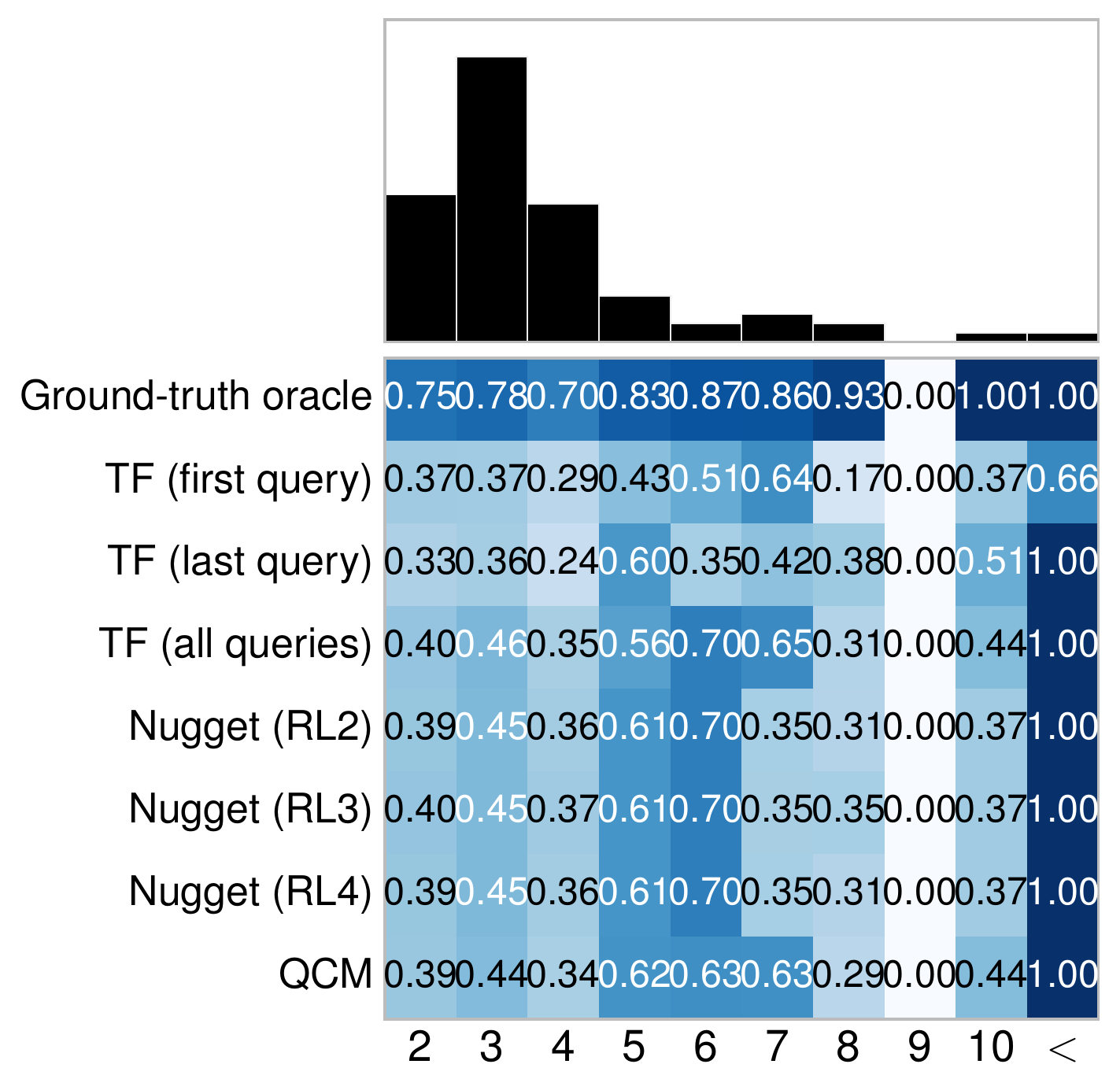}%
\end{subfigure}%
\hfill%
\lengthplot{2011}%
\hfill%
\lengthplot{2012}%
\hfill%
\lengthplot{2013}%
\hfill%
\lengthplot{2014}%
}

\caption{The top row depicts the distribution of session lengths for the 2011--2014 TREC Session tracks, while the bottom row shows the performance of the \TF{}, \Nugget{} and \QCM{} models for different session lengths.\label{fig:session_length}}

\vspace*{-0.75\baselineskip}
\end{figure*}

In Fig.~\ref{fig:session_length} we investigate the effect of varying session lengths in the session logs. The distribution of session lengths is shown in the top row of Fig.~\ref{fig:session_length}. For the 2011--2013 track editions, most sessions consisted of only two queries. The mode of the 2014 edition lies at 5 queries per session. If we examine the performance of the methods on a per-session length basis, we observe that the \TF{} methods perform well for short sessions. This does not come as a surprise, as for these sessions there is only a limited history that specialized methods can use. However, the \TF{} method using the concatenation of all queries still performs competitively for longer sessions. This can be explained by the fact that as queries are aggregated over time, a better representation of the user's information need is created. This aggregated representation naturally emphasizes important \emph{theme terms} of the session, which is a key component in the \QCM{} \citep{Yang2015}.

\begin{figure}[th!]

\begin{subfigure}[b]{\columnwidth}
	\includegraphics[width=\columnwidth]{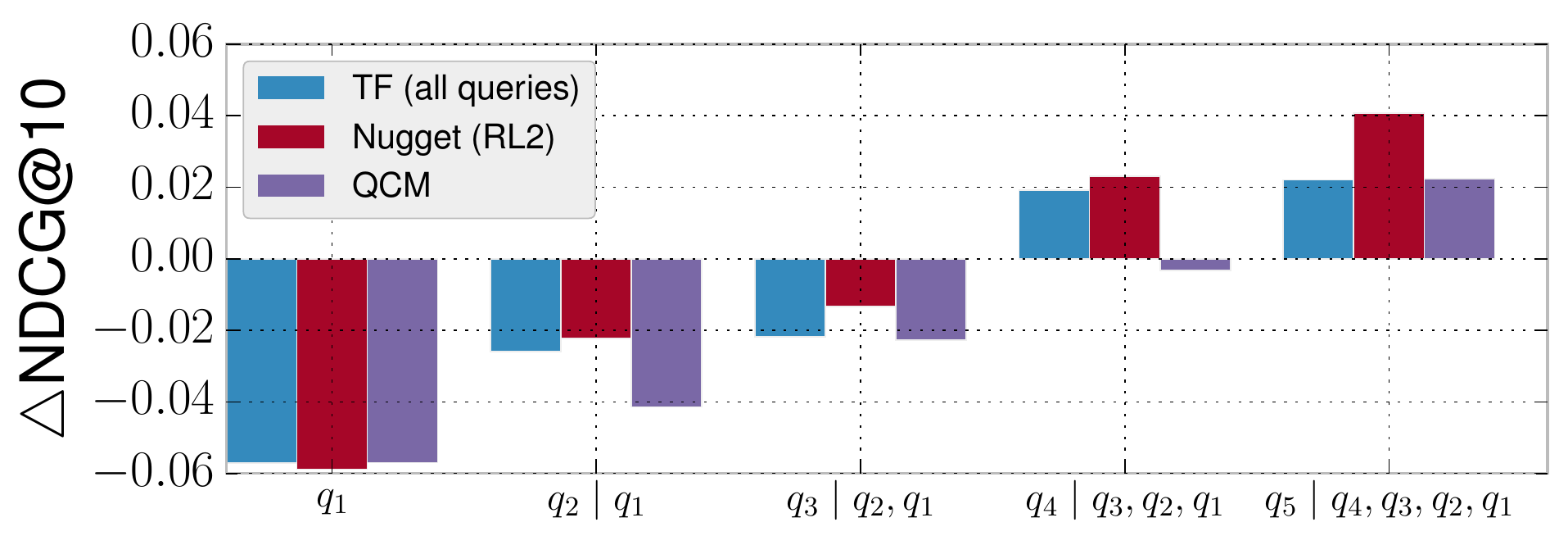}

	\caption{Full history of session\label{fig:progressing_session:all}}
\end{subfigure}

\begin{subfigure}[b]{\columnwidth}
	\includegraphics[width=\columnwidth]{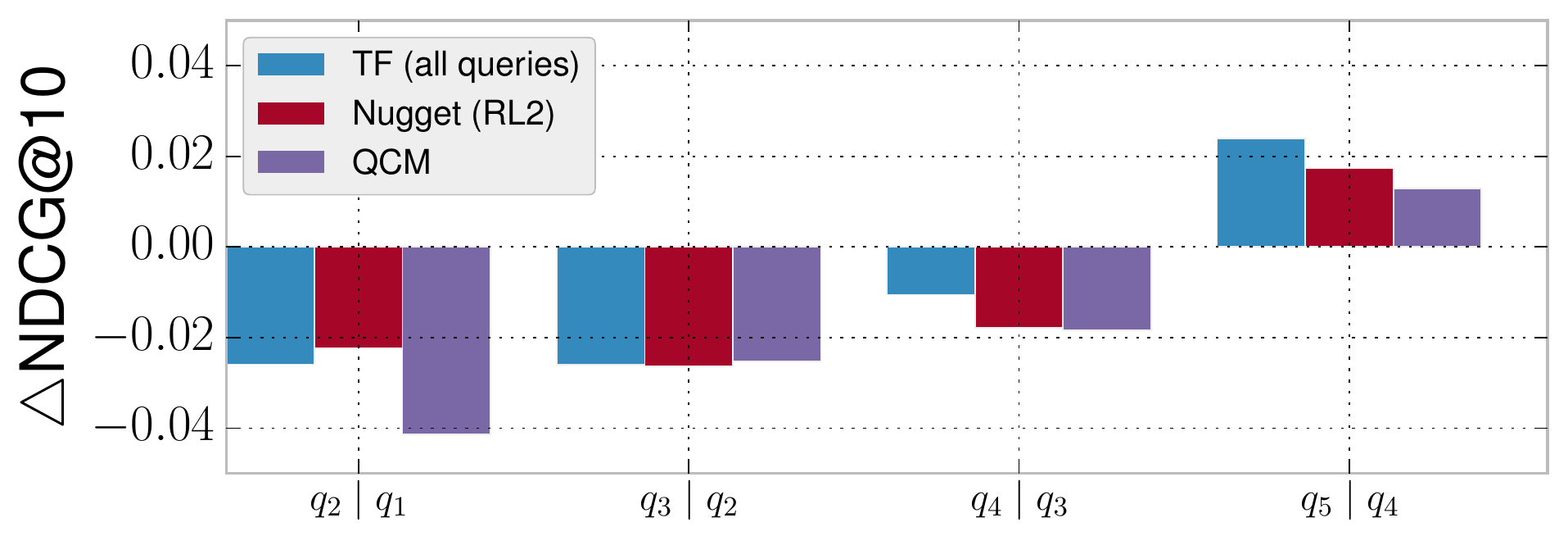}

	\caption{Previous query in session only\label{fig:progressing_session:previous}}
\end{subfigure}

\medskip

\caption{Difference in \NDCG{} with the official TREC baseline (\TF{} using the last query only) of $5$-query sessions (\numprint{45} instances) with different history configurations for the 2011--2014 TREC Session tracks.\label{fig:progressing_session}}

\vspace*{-0.75\baselineskip}

\end{figure}

\begin{table}[th!]
	\centering

	\caption{\NDCG{} for \TF{} weighting (\S\ref{section:methodology}), ideal term weighting (\S\ref{section:experiments:ideal}) and the ground-truth oracle (\S\ref{section:experiments:methods}).\label{tbl:lexical_results}}

	{\begin{tabular}{c@{ }c@{ }c@{ }c@{ }c}
\toprule
\multirow{2}{*}{} & \multicolumn{1}{c}{2011} & \multicolumn{1}{c}{2012} & \multicolumn{1}{c}{2013} & \multicolumn{1}{c}{2014} \\ 
\cmidrule(lr){2-2}
\cmidrule(lr){3-3}
\cmidrule(lr){4-4}
\cmidrule(lr){5-5}
\multicolumn{1}{l}{TF (all queries)} & $0.391$ & $0.333$ & $0.179$ & $0.183$ \\ 
\multicolumn{1}{l}{Ideal term weighing} & $0.589$ & $0.528$ & $0.361$ & $0.296$ \\ 
\multicolumn{1}{l}{Ground-truth oracle} & $0.716$ & $0.682$ & $0.593$ & $0.453$ \\ 
\bottomrule
\end{tabular}
}

	\vspace*{-0.75\baselineskip}
\end{table}

\break

How do these methods perform as the search session progresses? Fig.~\ref{fig:progressing_session} shows the performance of sessions of length five after every user interaction, when using all queries in a session (Fig.~\ref{fig:progressing_session:all}) and when using only the previous query (Fig.~\ref{fig:progressing_session:previous}). We can see that \NDCG{} increases as the session progresses for all methods. Beyond half of the session, the session search methods outperform retrieving according to the last query in the session. We see that, for longer sessions, specialized methods (\Nugget{}, \QCM{}) outperform generic term frequency models. This comes as no surprise. \citet{Bennett2012} note that users tend to reformulate and adapt their information needs based on observed results and this is essentially the observation upon which \QCM{} builds.

Fig.~\ref{fig:boxplot} and Table~\ref{tbl:results} reveal a large \NDCG{} gap between the compared methods and the ground-truth oracle. How can we bridge this gap? Table~\ref{tbl:lexical_results} shows a comparison between frequency-based term weighting, the ideal term weighting (\S\ref{section:experiments:ideal}) and the ground-truth oracle (\S\ref{section:experiments:methods}) for all sessions consisting of 7 unique terms or less (\S\ref{section:experiments:ideal}). Two important observations. There is still plenty of room for improvement using lexical query modeling only. Relatively speaking, around half of the gap between weighting according to term frequency and the ground-truth can be bridged by predicting better term weights. However, the other half of the performance gap cannot be bridged using lexical matching only, but instead requires a notion of semantic matching \cite{Li2014}.


\section{Conclusions}
\label{section:conclusions}

We have shown that naive frequency-based term weighting methods perform on par with specialized session search methods on the TREC Session track (2011--2014).\footnote{An open-source implementation of our testbed for evaluating session search is available at \paperImplementationUrl{}.} This is due to the fact that shorter sessions are more prominent in the session query logs. On longer sessions, specialized models are able to exploit session history more effectively. Future work should focus on creating benchmarks consisting of longer sessions with complex information needs.
Perhaps more importantly, we have looked at the viability of lexical query matching in session search. There is still much room for improvement by re-weighting query terms. However, the query/document mismatch is prevalent in session search and methods restricted to lexical query modeling face a very strict performance ceiling. Future work should focus on better lexical query models for session search, in addition to semantic matching and tracking the dynamics of contextualized semantics in search.
\vspace*{2mm}%
\begin{spacing}{-1}
\small
\noindent%
\textbf{Acknowledgments} This work was supported by the Google Faculty Research Award and the Bloomberg Research Grant programs. Any opinions, findings and conclusions or recommendations expressed in this material are the authors' and do not necessarily reflect those of the sponsors.
The authors would like to thank Daan Odijk, David Graus and the anonymous reviewers for their valuable comments and suggestions.
\end{spacing}%
\vspace*{3mm}%

\renewcommand{\bibsection}{
 \section*{REFERENCES}
}
\setlength{\bibsep}{-1pt}
\bibliographystyle{abbrvnatnourl}
{
\small
\raggedright
\vspace*{-1.0\baselineskip}
\bibliography{ictir2016-lexical-modeling}
}

\end{document}